# Segmentation of TCD Cerebral Blood Flow Velocity Recordings

*Federico Wadehn, Andrea Fanelli and Thomas Heldt*

*Abstract—* A binary beat-by-beat classification algorithm for cerebral blood flow velocity (CBFV) recordings based on amplitude, spectral and morphological features is presented. The classification difference between 15 manually and algorithmic annotated CBFV signals is around 5%.

## I. Introduction

In neurointensive care, cerebral blood flow velocity (CBFV) is a commonly recorded signal used for diagnosing and monitoring of various conditions such as vasospasms, arterial occlusions and ischemic strokes [1]. Recently CBFV, in conjunction with arterial blood pressure (ABP) has been used for non-invasive assessment of intracranial pressure [2]. Most commonly CBFV is recorded using transcranial Doppler (TCD), which estimates blood flow velocities in the main cerebral arteries based on the Doppler shift of the emitted MHz ultrasound signal. However, even in an intensive care setting, measuring cerebral blood flow velocity via TCD is a challenging task and results in strongly varying signal quality, due to uncertainty in the angle of insonation, movement of the sensor (especially in hand-held TCD) and unfavorable cranial anatomy [1]. Therefore, segmenting the CBFV recording (and the ABP signal) into good and bad segments is a crucial preprocessing step.

We will present a hierarchical, binary beat-by-beat classification algorithm for CBFV recordings based on the following three main steps: 1.) Hard thresholding 2.) Spectral correlation between ABP and CBFV signal and 3.) Self-similarity of CBFV waveform. Cerebral blood flow velocities from the *left/right middle cerebral artery (MCA)* and the *left/right posterior communicating artery (PCA)* were recorded from 8 subjects, for around 5 min each, using the Spencer ST3 handheld TCD.

F. W. is with the Swiss Federal Institute of Technology, Zurich, Switzerland (e-mail: wadehn@isi.ee.ethz.ch).

A. F. and T. H., are with the Massachusetts Institute of Technology, Cambridge, 02139 MA, USA (e-mail: fanelli@mit.edu, thomas@mit.edu).

## II. Binary Classification Algorithm

As a first step we perform a beat-onset detection using the *wabp* beat-onset algorithm from [3] adapted to CBFV by scaling the CBFV signal. On a beat-by-beat basis, each beat which features values below 5 cm/s and 300 cm/s is flagged as an artifact. Following this, CBFV signal segments labeled as good are assessed on a (non-overlapping) window-by-window basis with a window size of 8s. For this purpose, signal segments are labeled as artifacts if the correlation coefficient between the power spectral density, on a window of 0.5Hz-5Hz, of the CBFV and simultaneously recoded ABP was below 0.5. As a final step, a template of a CBFV pulse is obtained on the remaining good segments, by computing the median of all accepted beats. This template is finally used to assess the CBFV on a beat-by-beat basis and discard beats which have a large normalized mean squared error.

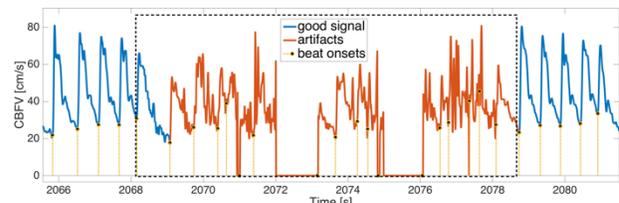

Fig. 1 shows a CBFV segment of a *posterior communicating artery* with both manual (black dotted box) and algorithmic labeling (red/blue).